\newcommand{\citebsm}{~\cite{Dev:2020drf,Wang:2025kqb,Rocha:2025rqc,Krnjaic:2024ols,Barman:2024lxy,Asai:2023mzl,Sieber:2023nkq,Atkinson:2022qnl,GrillidiCortona:2022kbq,Galon:2022xcl,Masiero:2020vxk}}
\newcommand{\citeyfs}{~\cite{Jadach:1996is,Jadach:1995nk,Schonherr:2008av,Krauss:2018djz,Hamilton:2006xz,Jadach:2001mp}}
\documentclass[a4paper,11pt]{article}
\pdfoutput=1
\usepackage{jheppub} 
\usepackage{lineno}
\usepackage{orcidlink}
\input{macros}
\usepackage{siunitx}
\usepackage{cleveref}
\begin{document}

\preprint{MCNET-25-30}

% \title{\textbf{Resummation of Soft Photons in \emupm} 
% }
\title{\textbf{Precise Predictions for \emupm at the MUonE Experiment }}
\abstract{
The proposed fixed-target experiment, MUonE, at CERN will aim to 
measure the hadronic contribution to the running of the QED 
coupling by analysing the scattering of muons on electrons. 
Here we present state-of-the-art predictions for the process 
\emupm, where for the first time an all-order resummation of soft 
and soft-collinear logarithms has been performed. Further, we match 
this resummation with the complete next-to-leading and the dominant 
next-to-next-to-leading higher-order corrections. We find that the 
resummation has a dominant effect in the signal region, while the 
systematic matching significantly reduces the perturbative uncertainty. 
}

\author{Alan~Price\orcidlink{https://orcid.org/0000-0002-0372-1060}}
\emailAdd{alan.price@uj.edu.pl}
\affiliation{Jagiellonian University, ul.\ prof.\ Stanis\l{}awa \L{}ojasiewicza 11, 30-348 Krak\'{o}w, Poland}
% \affil{Jagiellonian University, ul.\ prof.\ Stanis\l{}awa \L{}ojasiewicza 11, 30-348 Krak\'{o}w, Poland}
\maketitle
\tableofcontents
\section{Introduction}
\label{sec:intro}

The primary goal of the proposed MUonE experiment~\cite{MUonE:LoI} at CERN will be to measure the hadronic contribution to
the electromagnetic coupling, \DeltaHad, from the elastic collisions of a high-intensity muon beam upon a fixed target to induce \emupm scattering. 
% With this measurement the leading order hadronic contribution to the muon magnetic moment, \ahlo,
% can be extracted.
Since the structure of this scattering process contains no resonance structures, like those present at low-energy \ee machines,  it creates an optimal
experimental environment for the extraction of \DeltaHad as proposed in~\cite{CarloniCalame:2015obs,Abbiendi:2016xup} in the space-like region. For this result to be competitive with previous measurements of \DeltaHad, such as ones from dispersion relations~\cite{Aliberti:2025beg,Davier:2019can,Keshavarzi:2018mgv,Jegerlehner:2018zrj,Jegerlehner:2006ju}
or lattice calculations~\cite{DellaMorte:2017dyu,FermilabLattice:2017wgj,Borsanyi:2017zdw,Meyer:2018til,Blum:2018mom,Giusti:2018mdh,Giusti:2019xct,Giusti:2019hkz,Shintani:2019wai,FermilabLattice:2019ugu,Gerardin:2019rua,Aubin:2019usy}, the uncertainty of the measured differential cross-sections at MUonE will have to be of the order of 10~ppm.
In addition to the extraction of \DeltaHad, the MUonE experiment will also be able to probe a range of exotic 
beyond the Standard Model (BSM) scenarios\citebsm.

For decades there has existed a discrepancy between the theoretical~\cite{Aoyama:2020ynm} and experimental~\cite{Muong-2:2021ojo,Muong-2:2023cdq} values of $a_\mu$~\cite{Jegerlehner:2009ry} leading to a plethora of studies~\cite{Athron:2025ets} which probed the physical implications of such an anomaly.
However, recently the Muon g-2 experiment~\cite{Muong-2:2015xgu} E989 at Fermilab, which has combined two results from data 
taken over the past few years with a previous result from the Brookhaven
National Laboratory (BNL) experiment~\cite{Muong-2:2006rrc} to reach an unprecedented accuracy of 0.20~ppm, has 
resolved this long standing anomaly with the uncertainties of the theory predictions. While this anomaly has been resolved within the Standard Model(SM),
there exists a tension between the measured HVP contribution which are of the level of 2.5-5$\sigma$. In particular,
there exists a strong discrepancy between the CMD-3 results~\cite{CMD-3:2023alj} from all previous measurements,
and the deviation was significantly large enough that in the recent update of SM predictions
the CMD-3 value could not be included in the common average of data-driven estimates for \ahlo.
In addition, an accurate measurement of \DeltaHad will be an important input for future \ee colliders~\cite{ILC:2013jhg,Aicheler:2012bya,FCC:2018evy,CEPCStudyGroup:2018ghi,Vernieri:2022fae},
in particular for luminosity measurements where the uncertainty on the HVP will need to be reduced~\cite{Jadach:2018jjo,Jegerlehner:2019lxt}, for which a direct measurement of \DeltaHad will be invaluable.  

On the theoretical side, \emupm scattering has been studied in the past~\cite{Nikishov:1961abc,Eriksson:1961abc,Eriksson:1963def,VanNieuwenhuizen:1971yn,Kukhto:1987uj,Bardin:1997nc,Alacevich:2018vez,Engel:2019nfw,CarloniCalame:2019mbo}
using various different approaches and recently substantial progress has been made in developing modern Monte-Carlo tools, such as McMule~\cite{Banerjee:2020rww} and MESMER~\cite{Alacevich:2018vez,CarloniCalame:2020yoz,Budassi:2021twh}, to compute electron–muon scattering at fixed-order accuracy at next-to-leading (\NLO) and next-to-next-to-leading (\NNLO) order and in which a wealth of recent theory calculations~\cite{Banerjee:2020tdt,Mastrolia:2017pfy,DiVita:2018nnh,Budassi:2021twh,Fael:2018dmz,Delto:2023kqv,Badger:2023xtl,Fael:2019nsf,Ahmed:2023htp,Fael:2022rgm,Mastrolia:2003yz,Fael:2022miw,Broggio:2022htr,Plestid:2024xzh,Bonciani:2021okt,Dave:2024ewl,Plestid:2024jqm,Engel:2023ifn,Engel:2023rxp,Gurgone:2024xdt,Balzani:2021del,Ronca:2019kcw,Engel:2018fsb} have been developed for.
However, despite these advances in the fixed-order calculations, considerably less attention has been given to another essential component of high-precision predictions, the resummation of soft and collinear logarithms arising from the emission of multiple photons. Depending on the kinematics of the phase space, these logarithms can become large and jeopardize the reliability of the perturbative expansion.
Indeed, it has already been shown in~\cite{Alacevich:2018vez,CarloniCalame:2020yoz,Broggio:2022htr}, that in the signal region, which corresponds to low scattering angle of the electron, the effects of infrared (IR) logarithms, associated 
with the emission of soft photons, becomes quite sizeable. 

One framework in which we can resum and tame the effects of such large logarithms is based on the Yennie–Frautschi–Suura (YFS) theorem~\cite{Yennie:1961ad}. This theorem allows us to reorganizes the entire perturbative expansion in such a way that all logarithms associated with infrared divergences are resummed to infinite order, leaving a finite remainder that can be computed to a high precision. It is also highly suitable for event generation, where it not only provides an accurate cross-section prediction, but in addition can generate the complete multi-photon~\cite{Jadach:1988gb} phasespace. The YFS theorem has been implemented in many event generators\citeyfs, where most have focused on specific processes or decays, while the first automation 
to arbitrary lepton scattering processes was presented in \Sherpa~\cite{Sherpa:2024mfk,Krauss:2022ajk}. Recently, we have expanded this implementation in \Sherpa to automatically include \NLOEW
and \NNLOEW corrections~\cite{Price:2025fiu}. It is this implementation that we will use in this paper to calculate precise theory predictions for the MUonE experiment, where will have
resummed all soft logarithms to infinite order, and additionally will provide the matching to the full \NLO corrections and we will in addition include the \NNLO corrections, where the only 
approximation we use will be in the two-loop corrections, however the dominant IR contributions will be included. Alternatively, one could consider resumming the large logarithms using a parton shower method~\cite{Balossini:2006wc,Flower:2022iew}, which may be explored in a future work.

\section{Theory}
\label{sec:theory}
\begin{figure}
    \centering
    \includegraphics[width=0.45\textwidth]{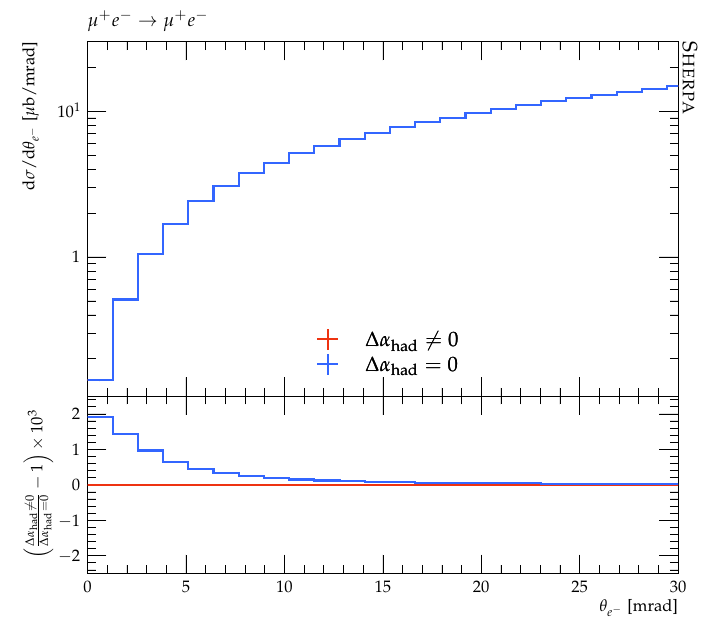}
    \includegraphics[width=0.45\textwidth]{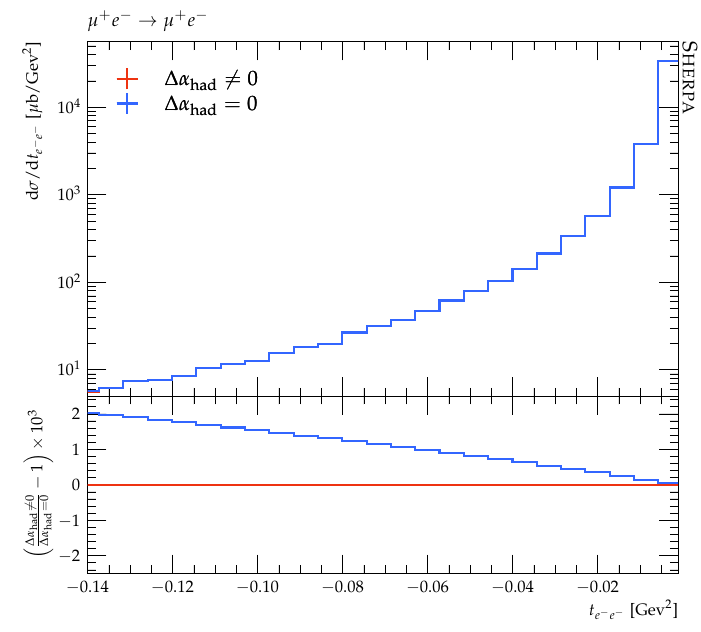}
    \caption{The impact of \DeltaHad on electron observables at the MUonE experiment. }
    \label{fig:lo-had}
\end{figure}    
In~\cref{fig:lo-had}, we show the impact that the HVP has to the \emup observables that will be measured at MUonE. We use a dedicated interface in \Sherpa to the stand-alone Fortran library \emph{hadr5x23}~\cite{Jegerlehner:1985gq,Jegerlehner:2006ju,Burkhardt:1989ky,Eidelman:1995ny,Jegerlehner:2003rx} to provide numerical values of \DeltaHad . We see that the largest effect can be seen at low electron angle, $\theta_e$, in particular below 5 \mrad where the effect is of order $10^{-3}$. In the right plot of~\cref{fig:lo-had} we see how the contribution depends on the Mandelstam variable \tee. The MUonE experiment will aim to extract \DeltaHad from this shape in the differential distributions by using a template fit method~\cite{MUonE:LoI}. Schematically, this can be thought as,
\begin{equation}
    \frac{d\sigma^{\textbf{MUonE}}\left(t\right)}{d\sigma^{\textbf{MC}}\left(t\right)\big|_{\DeltaHad=0}} \approx 1+2\DeltaHad
\end{equation}
where $d\sigma^{\textbf{MUonE}}\left(t\right)$ will be the measured differential distributions and 
$d\sigma^{\textbf{MC}}\left(t\right)$ are the MC predictions, where the numerical effects of \DeltaHad are neglected. Therefore, it is crucial that any MC generator provides accurate predictions for the differential distributions, and it is this contribution that we will focus on in the rest of this work.

The YFS theorem states that the differential cross-section, encompassing all possible real and virtual photon emissions in an underlying Born process, is given by,
\begin{align}\label{eq:masterYFS}
\ddone\sigma^{(\infty)} &= 
\sum_{n_\gamma=0}^\infty \frac{e^{Y(\Omega)}}{n_\gamma!}\,
\ddone{\Phi_n}
\left[\prod_{i=1}^{n_\gamma}\ddone{\Phi_i^\gamma}\,\eik{i}\,\Theta(k_i,\Omega)\right]
\left(\tilde{\beta}_0(\Phi_n)
+\sum_{j=1}^{n_\gamma}\frac{\tilde{\beta}_1(\Phi_{n+1})}{\eik{j}}
\right. \notag \\[4pt]
&\qquad\left.
+\sum_{\substack{j,k=1\\ j<k}}^{n_\gamma}
\frac{\tilde{\beta}_{2}(\Phi_{n+2})}{\eik{j}\eik{k}}
+ \cdots\right)\,.
\end{align} 
Here, $\done\Phi_n$ denotes the modified final state phase space element, 
the $\done\Phi_i^\gamma$ are the phase space elements spanned by the $n_\gamma$ 
real photon momenta $k_i$ emitted off the leading order configuration. 
The infrared (IR) finite $\tilde{\beta}$ are the process-dependent terms in which we can include higher-order corrections.
In the above expression, we have suppressed the indexing of unresolved emissions by introducing the following notation,
\begin{align}
\tilde{\beta}_{n_R} = \sum_{n_V=0}^\infty\tilde{\beta}_{n_R}^{n_V+n_R}\,,
\end{align}
where $n_R$,$n_V$ are the number of resolved and unresolved emissions under consideration. The \NLOEW corrections can be included 
by considering both the virtual, \betatilde{0}{1} and real, \betatilde{1}{1}, corrections which can be calculated by,
\begin{align}
    \betatilde{0}{1}\left(\Phi_n\right) =& \mathcal{V}(\Phi_n) - \sum_{ij}\mathcal{D}_{ij}\left(\Phi_{ij}\otimes\Phi_n\right)\label{EQ:Virtual},\\
     \betatilde{1}{1}\left(\Phi_{n+1}\right)  =& \frac{1}{2\left(2\pi\right)^3}\mathcal{R}\left(\Phi_{n+1}\right) 
                                                - \sum_{ij} \mathcal{\tilde{D}}_{ij}\left(\Phi_{ij}\otimes\Phi_k\right). \label{EQ:Real}
\end{align}
Here $\mathcal{R}\left(\Phi_{n+1}\right)$ represents the squared amplitude which arises from all diagrams that can be constructed by considering one additional real photon to the born process and 
similarly $\mathcal{V}(\Phi_n)$ represents the complete virtual \NLO corrections. The second terms in~\cref{EQ:Virtual,EQ:Real} are the corresponding subtraction terms which render the 
correction IR finite, which are automatically calculated in \Sherpa~\cite{Price:2025fiu}.  At NNLO, we need to include additional corrections, namely,
\begin{align}
\betatilde{1}{2}(\Phi_{n+1}) &= \mathcal{RV}(\Phi_{n+1})
  - \sum_{ij}\mathcal{D}^{(1)}_{ij}(\Phi_{ij+1}\otimes\Phi_n)\label{EQ:RV}, \\[0.5em]
\tilde{\beta}_2^2(\Phi_{n+2}) &= \mathcal{RR}(\Phi_{n+2})
  - \eik{1}\betatilde{1}{1}(\Phi_{n+1};k_2)
  - \eik{2}\betatilde{1}{1}(\Phi_{n+1};k_1) \nonumber\\
&\quad  - \eik{1}\eik{2}\betatilde{0}{0}(\Phi_n),\label{EQ:RR} \\[0.5em]
\betatilde{0}{2}(\Phi_n) &= \mathcal{VV}(\Phi_n)
  - \sum_{ij}\mathcal{D}_{ij}(\Phi_{ij}\otimes\Phi_n)\,\betatilde{0}{1}(\Phi_n)\nonumber \\
 &\quad - \frac{1}{2}\left(\sum_{ij}\mathcal{D}_{ij}(\Phi_{ij}\otimes\Phi_n)\right)^2
  \betatilde{0}{0}\left(\Phi_n\right),
\label{EQ:VV}
\end{align}
where $\mathcal{RV}(\Phi_{n+1})$ is the real-virtual, $\mathcal{RR}\left(\Phi_{n+2}\right)$ is the double real, and $\mathcal{VV}(\Phi_n)$ is the double virtual corrections, and 
again each contribution has its own subtraction term. The explicit forms of the subtraction terms for each contribution can be found in~\cite{Price:2025fiu}.
Currently, there is no publicly available calculation of the two-loop contribution that suits our needs. In particular, we need a calculation that takes into account all mass terms explicitly such that the collinear divergences can be regulated. Although there has been some progress in the calculation of such amplitudes for other processes, such as Bhabha~\cite{Delto:2023kqv}, the hierarchy of masses in our process remains a challenging bottleneck.  
We stress that when such a calculation becomes available for us, it can be easily incorporated 
into our framework. For now we shall only include the dominant IR contributions from the double-virtual while the sub-leading non-IR enhanced corrections are neglected, which is a similar approach as~\cite{CarloniCalame:2020yoz}. 

\begin{figure}
    \centering
    \includegraphics[width=0.45\linewidth]{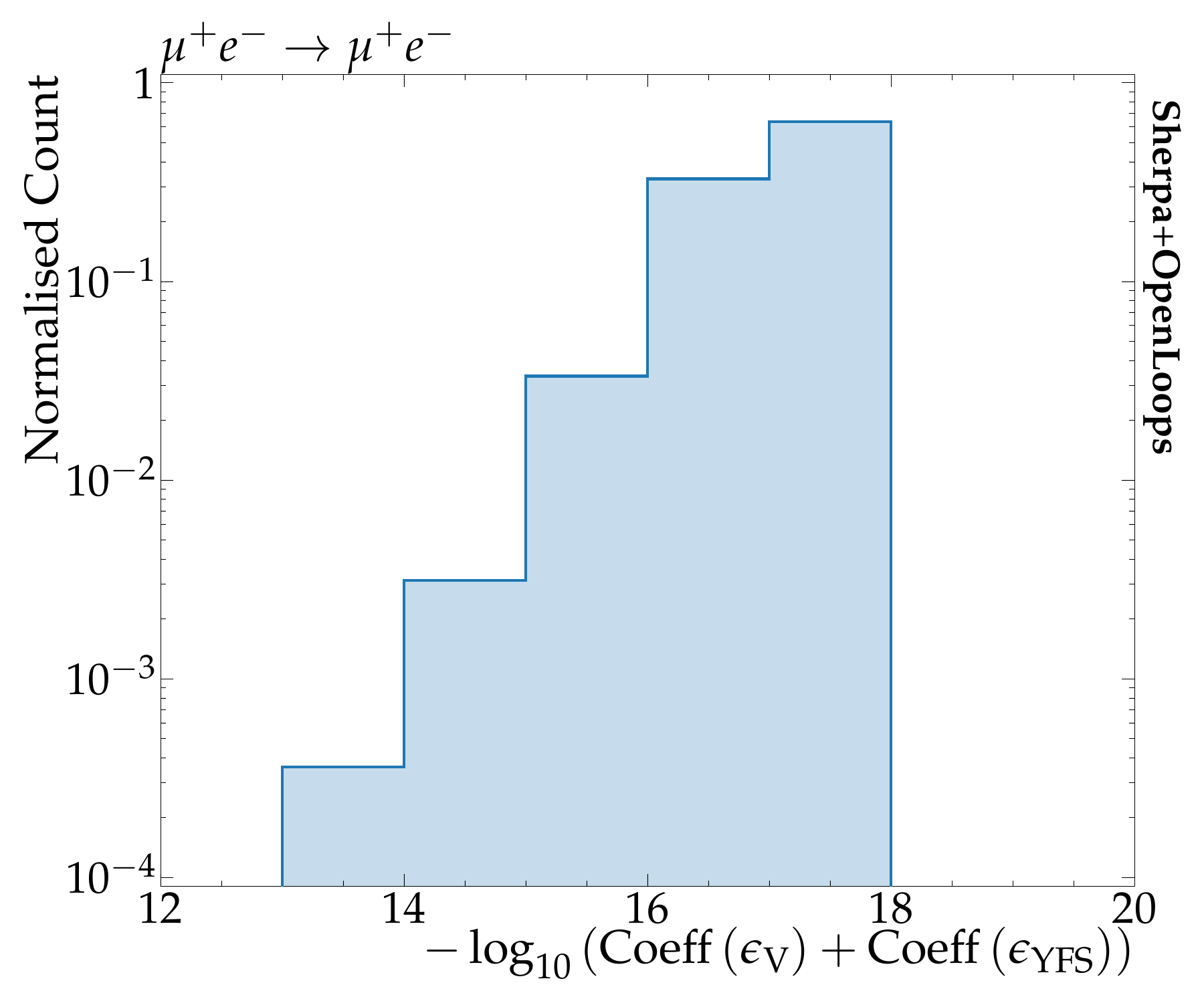}
    \includegraphics[width=0.45\linewidth]{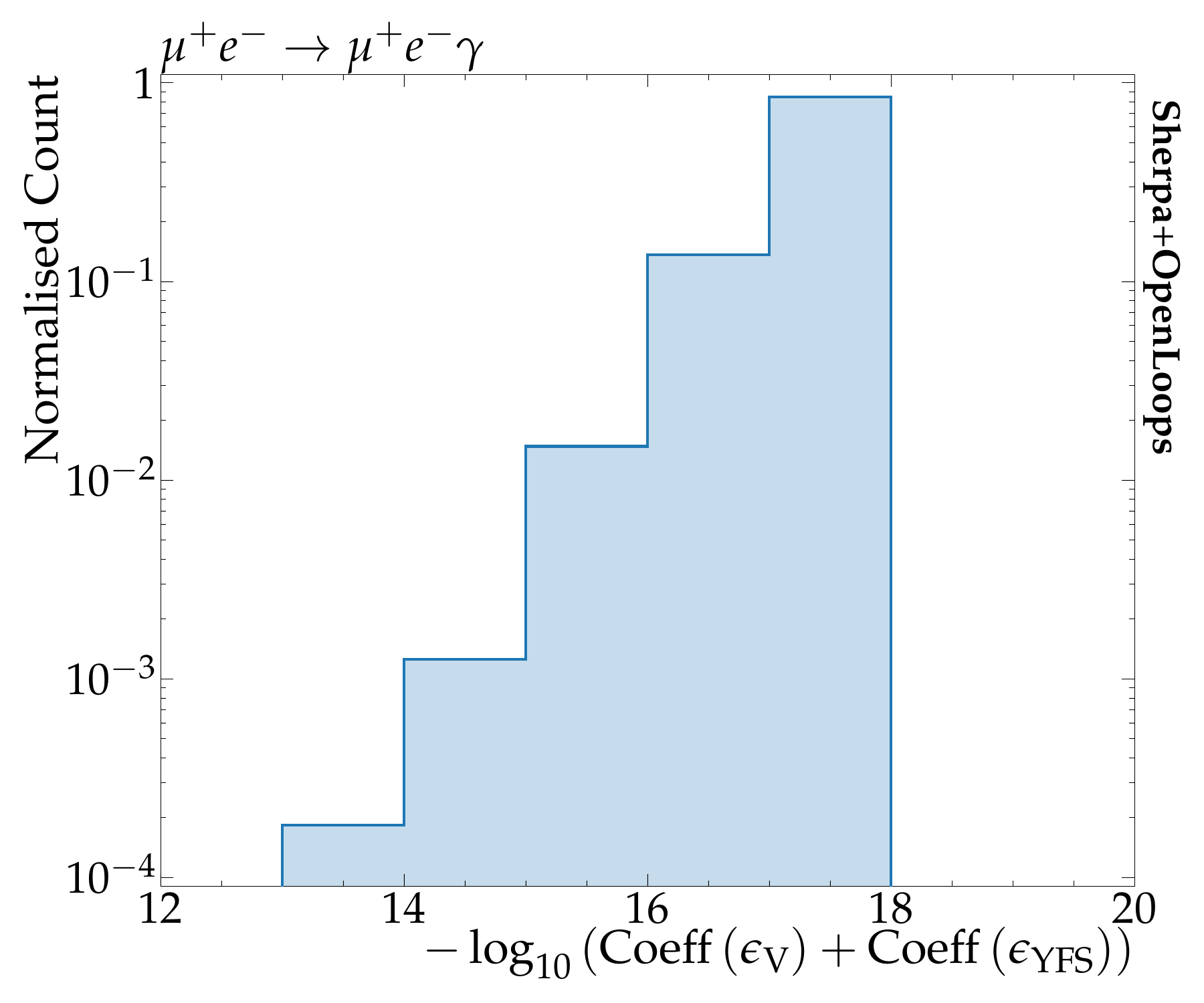}
    \includegraphics[width=0.45\linewidth]{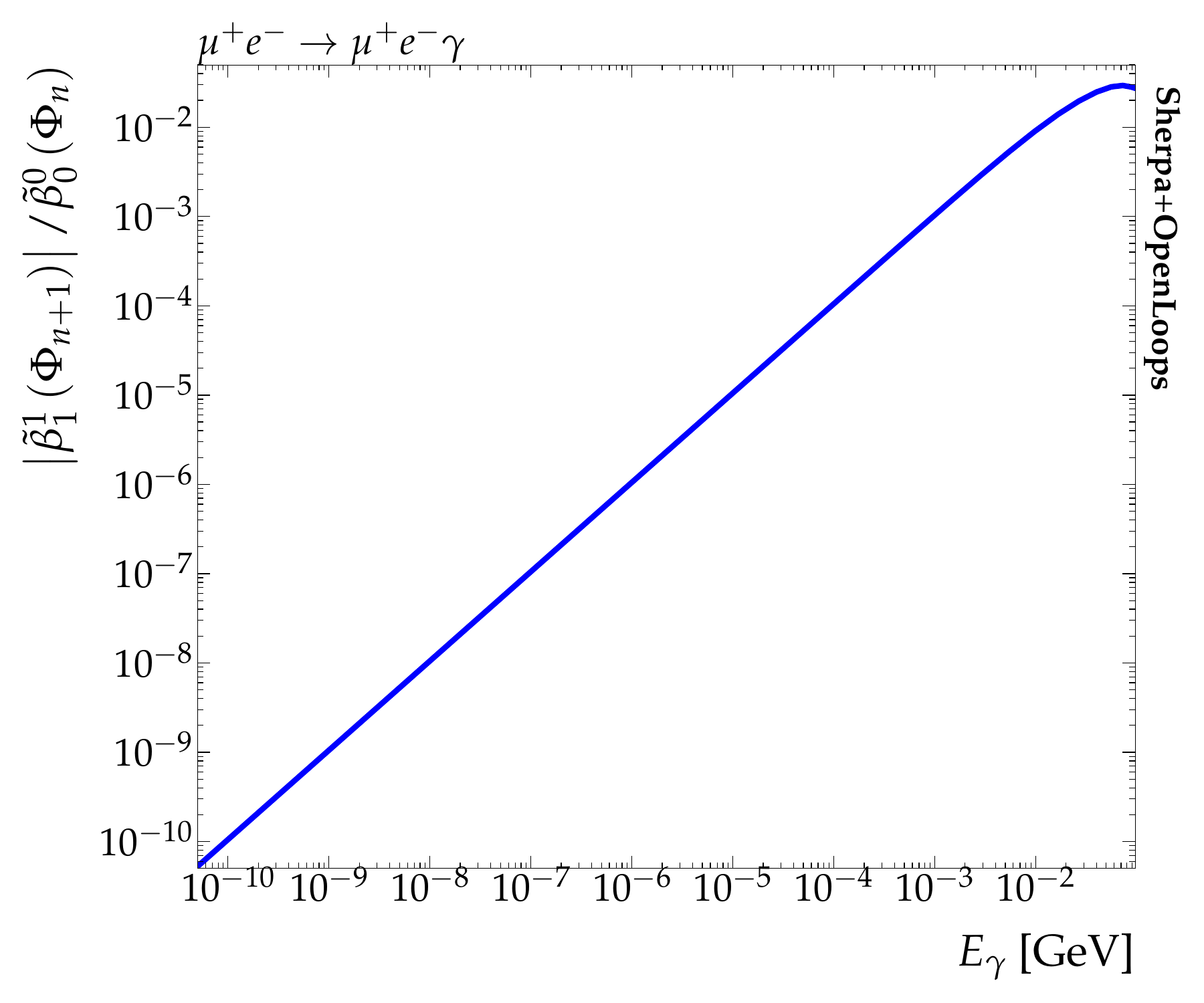}
    \includegraphics[width=0.45\linewidth]{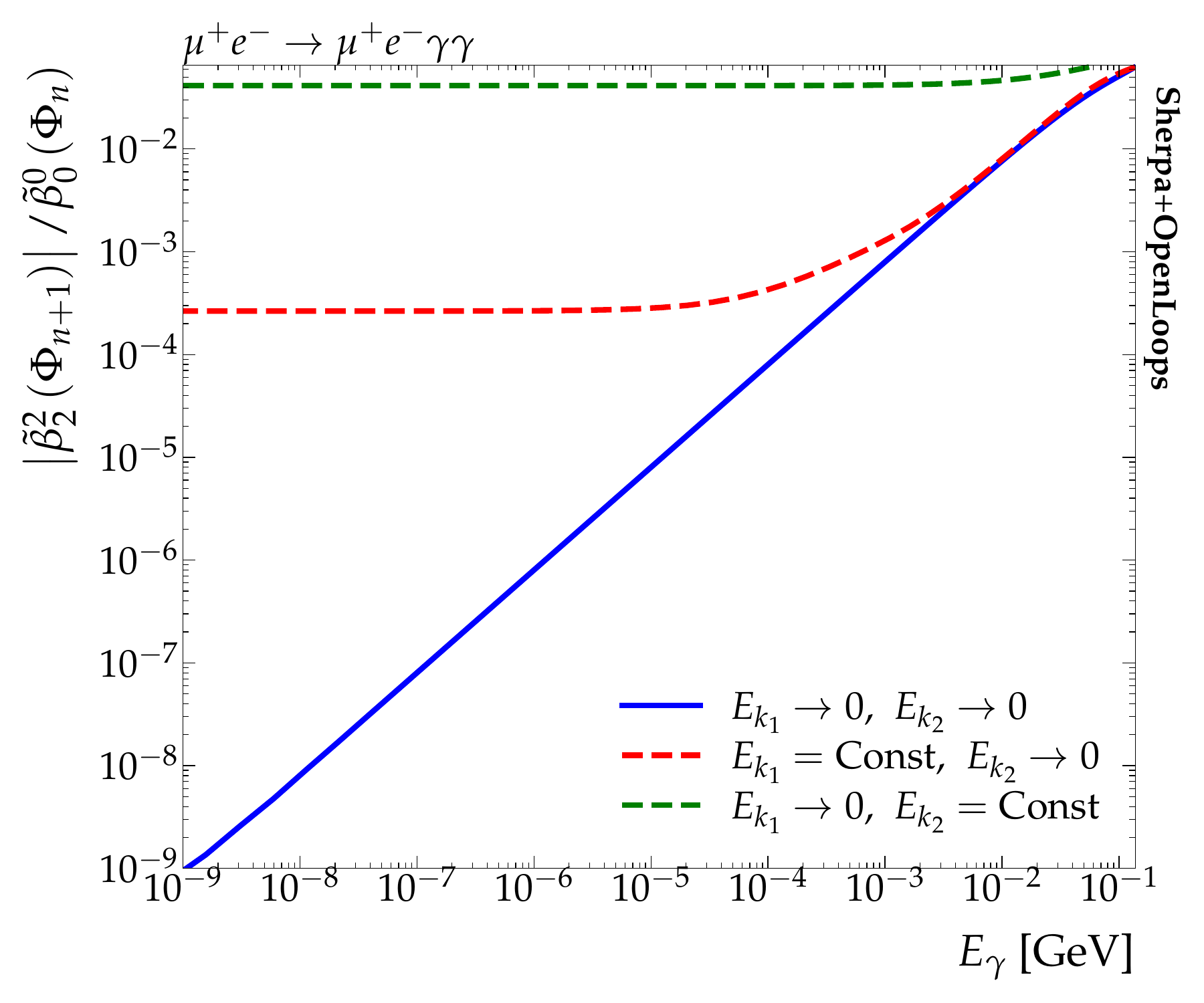}
    \caption{Top: The number of digits the pole cancellations are achieved to for 
            both the virtual (left) and real-virtual (right) corrections
            in dimensional regularization.
            Bottom: The behaviour of the real (left) and double-real (right) corrections in 
            the limit where photon momentum becomes ultra-soft.}
    \label{fig:IRPlots}
\end{figure}

The tree-level amplitudes are obtained using one of \Sherpa's internal automated matrix-element generators, 
either \Amegic~\cite{Krauss:2001iv} or \Comix~\cite{Gleisberg:2008fv}. The one-loop corrections are computed 
through dedicated interfaces~\cite{Biedermann:2017yoi,Kallweit:2015fta} to external tools such as \Recola~\cite{Actis:2016mpe} and \OpenLoops~\cite{Buccioni:2019sur} which in turn use tools such as \Collier~\cite{Denner:2016kdg}, \CutTools~\cite{Ossola:2007ax} and \OneLoop~\cite{vanHameren:2010cp}
to help their numerical evaluations. 
We demonstrate in~\cref{fig:IRPlots} how the subtraction procedure successfully removes the IR-divergent contributions.
In the upper row, we illustrate how the $\epsilon$ poles that encode the infrared divergences of the amplitudes 
in a loop cancel against the poles in the YFS subtraction term. For most events, this cancellation is accurate to 15 digits or more, which we consider sufficient for both the virtual and the real-virtual corrections.
The bottom row shows the behaviour of the tree-level contributions for the real emissions  coming from~\cref{EQ:Real,EQ:RR}, normalised to the Born. On the left, we display the behaviour of~\cref{EQ:Real} 
in the limit where the photon energy becomes soft. This limit is especially relevant for MuonE, since in the signal region used for 
extracting~\DeltaHad, the electron phasespace is constrained such that soft-photon radiation becomes dominant. 
Consequently, achieving precise predictions requires a matching procedure that remains numerically stable in 
the soft-emission limit, as confirmed by the lower-left panel of~\cref{fig:IRPlots}. We observe that the contribution 
to~\cref{EQ:Real} becomes negligible as the photon energy approaches zero, reflecting the fact that the effects of such emissions are already incorporated in the resummation. In the bottom right of~\cref{fig:IRPlots} we examine the limiting behaviour of~\cref{EQ:RR}. For such contributions, there are three limits of interest, namely the limit where one photon becomes soft and the other remains hard, and the limit where both photons become soft.
In the former case, we expect the hard photon contribution will saturate the contribution while the soft photon contribution is removed by the subtraction term. In the case where both photons become soft, we expect the overall contribution to approach 0, which we see in our plot.

\begin{figure}
\centering
    \includegraphics[width=0.65\linewidth]{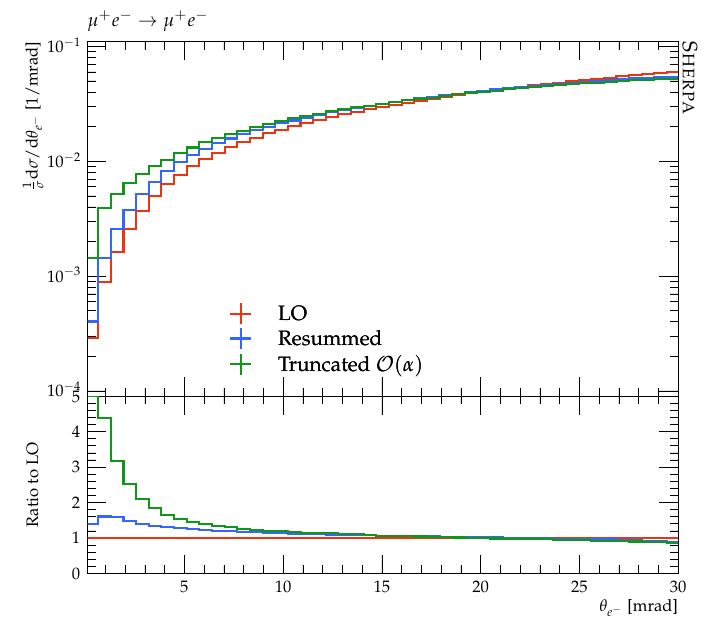}
    \caption{Leading-order, resummed, and truncated predictions for the electron's scattering angle.
            The truncated expansion includes only the  \NLO fixed-order corrections. }
\label{fig:Resum}
\end{figure}
 
Before we discuss the full physical results of this paper, it is worthwhile examining the effect of our 
resummation in the signal region of the MUonE experiment. It is possible to undo the YFS resummation to investigate
its effects, and this can be achieved
by expanding~\cref{eq:masterYFS} and truncating it to a given fixed order within the MC. This truncated expression
can then be used as an approximation of the fixed-order corrections where effects of resummation have been removed. And while the full resummed results will include multiple photon emission, the truncated approximation will only include a fixed number of photons.
In~\cref{fig:Resum}, we present a prediction where we truncate the YFS expansion at \oforder{\alpha} and include both the leading-order (LO) and resummed predictions. At large scattering angles, there is little difference between the resummed and truncated predictions. However, in the signal region, $\theta_e \leq 5$, a significant difference emerges. Here, the photon emissions from the electron line are quite soft, leading to large logarithmic corrections that increase as we approach $\theta_e \rightarrow 0$. By enabling YFS, we can tame this divergent behaviour by resumming the problematic logarithms. While there’s still a correction of approximately 50\% compared to the LO predictions, we’ll show in the next section that this is purely a perturbative uncertainty that can be systematically improved by including higher-order corrections. 
% We have observed similar size and shape distributions from the fixed-order Monte-Carlo simulations, McMule, and MESMER, when comparing the ratios of NLO to LO as our truncated prediction. 
This indicates that the fixed-order expansion is unreliable in this region and to have physically sensible results for the MUonE experiment it will be mandatory to include resummation.

\section{Results}
\label{sec:results}
%!TEX root = main.tex
% \subsection{Setup}
For real-world studies of \emupm we will consider both the complete one-loop and real corrections in association with the full resummation of both ISR and FSR which will allow us to provided matched \NLOEW predictions. In addition, we will examine the effects of including the double real and real-virtual corrections, while the full two loop corrections
are approximated using the YFS method similar to~\cite{CarloniCalame:2020yoz}.
All leptons are treated as fully massive and the multiphoton phasespace~\cite{Krauss:2022ajk} will be treated completely analytically.
\begin{table}
\setlength{\tabcolsep}{11pt}
\renewcommand{\arraystretch}{1.5}
  \begin{center}
    \begin{tabular}{c| c | c}
      & Mass [ \UGeV] & Width [ \UGeV] \\
      \hline
      \hline
         Z & 91.1876 & 2.4952 \\
         W & 80.398 & 2.085\\
         e & 0.000511  & - \\
     $\mu$ & 0.105 & - \\
     \hline
     \hline
     $\alpha^{-1}\left( 0 \right)$ & 137.03599976
    \end{tabular}
    \parbox{0.8\textwidth}{\caption{\label{tab:Validation:ew-inputs} 
        Electroweak input parameters in the $\alpha(0)$ scheme.}}
  \end{center}
\end{table}
Throughout this calculation, we employ the $\alpha(0)$ scheme where the input parameters are $\alpha$, $M_W$, and $M_Z$ and the explicit values can be found in~\cref{tab:Validation:ew-inputs}. We do not consider any of the background processes, which have been studied in great detail in~\cite{Abbiendi:2024swt,Budassi:2022kqs,Budassi:2021twh}. We do note that in \Sherpa it is possible to simulate $\mu^{\pm}e^-\rightarrow\mu^{\pm}e^-\ell \bar{\ell}$ within our YFS framework but as we identify this as a separate unique LO process we will not address it here but defer it to a future work.
In the lab frame, we assume the muon beam is travelling in the positive z-direction with an energy of 150~\UGeV~\footnote{While the actual beam will be 160~\UGeV, this value was chosen to align with previous studies by event generators. This has no effect on our conclusions.}. 
We have implemented a dedicated fixed-target mode in \Sherpa where the initial beams will be given as,
\bea 
p^{\text{Lab}}_{\mu^\pm} &=& \left(150,0,0,p^z_{\mu^\pm}\right)\nonumber\\
p^{\text{Lab}}_{e^-} &=& \left(m_e,0,0,0\right).
\eea
We then boost the beams to the centre-of-mass frame, where we perform our calculation, and then finally boost back to lab frame where we analyse our events with the \Rivet package~\cite{Bierlich:2024vqo,Buckley:2010ar}. 
For all our scenarios we follow a similar setup compared to previous fixed-order studies~\cite{CarloniCalame:2020yoz} and 
apply the following phasespace cuts in the lab frame,
\bea
E_{e^-} &>& 1~\UGeV\,\nnb \\
\theta_e,\theta_\mu &<& 100~\mrad.
% \left|\pi - \left|\phi_e - \phi_\mu\right|\right| &<& 3.5 \text{ mrad}.
\eea
and additionally we also include two separate cuts on the acoplanarity,
\begin{equation}
\left|\pi - |\Delta \phi_{\mu^+e^-}|\right| \leq \left[3.5 ~\mrad,0.4~\rad \right]
\end{equation}
where the first cut follows from previous studies 
while the second is a more realistic experimental cut for the MUonE experiment~\cite{Spedicato:2025qid,Abbiendi:2024swt}. 
A strict cut, such as our first choice, will remove a significant amount of radiative events and enhance the number of 
elastic events. 
It is a useful cut for examine the impact of soft radiation which is 
not completely removed by such a cut. 
% Since our YFS is fully exclusive with respect to photon multiplicity, we employ a dressing algorithm when analysing events. This algorithm clusters the lepton’s four-momentum with nearby photons within a predetermined dressing cone. For this study, we present results with a dressing cone size of $\Delta R = 0.4$. Photons which have been clustered are subsequently removed from the event record.

We will consider three levels of correction, \yfslo, \yfsnlo, and \yfsnnlo. The \yfslo predictions will include the 
resummation of both the initial and final photon emissions, including interference effects, while the perturbative 
expansion will be truncated to LO only, which is simply the \betatilde{0}{0} term.
The \yfsnlo distributions include 
the same resummation, however the complete resolved and unresolved \oforder{\alpha} corrections are included 
as described in~\cref{EQ:Real,EQ:Virtual}.
Finally, the \yfsnnlo predictions will additionally include the complete double real and real-virtual corrections,
while we employ a YFS inspired approximation to the two-loop, where the dominant IR corrections have be resummed 
while the sub-leading non-IR enhancements have been included in the leading-log approximation~\cite{Jadach:2000ir}.
Due to the missing non-IR terms in the two-loop corrections we use a lower-case "\emph{n}" to represent these corrections.
All one-loop amplitudes which appear in \betatilde{0}{1} and \betatilde{1}{2} will be calculated using \Sherpa's interface
to \OpenLoops to calculate the virtual corrections. The remaining tree-level amplitudes, subtraction terms, and phasespace integrations 
are handled automatically by \Sherpa. Similarly to~\cite{CarloniCalame:2019mbo} we find the weak contributions have small 
effect and neglect them in the remainder of this work.

We will calculate the size of our corrections by taking the difference with the next lowest order prediction,
\begin{equation}\label{EQ:Uncert}
    \Delta \mathrm{YFS}_{i} = \left(\frac{\mathrm{YFS}_{i}}{\mathrm{YFS}_{i-1}}-1\right)\times 100 \quad\quad  i\in (\text{LO},\text{NLO},\text{nNLO})
\end{equation}
and for \Deltayfs{LO} we take the reference to be the Born-level prediction without any resummation effects.

% \begin{table}[h!]
% \centering
% \begin{tabular}{lccc}
% \hline\hline
% \textbf{Process} & \yfslo \Mpb & \yfsnlo \Mpb & \yfsnnlo \Mpb  \\
% \hline
% \emup & $246.452(1)$ & $251.156(4)$ & $252.03(5)$ \\
% \emum  & $246.451(1)$ & $252.19(1)$ & $253.05(2)$ \\
% \hline\hline
% $\left|\pi - \bigl|\Delta \phi_{\mu^+e^-}\bigr|\right| \leq 3.5  \mrad$ &  &  & \\
% \hline
% \emup & $246.452(1)$ & $251.156(4)$ & $252.03(5)$ \\
% \emum  & $246.451(1)$ & $252.19(1)$ & $253.05(2)$ \\
% \end{tabular}
% \caption{Cross sections for Process A and Process B at different perturbative orders.}
% \end{table}

% \begin{table}[h!]
% \centering
% \begin{tabular}{l S[table-format=3.3(2)] S[table-format=3.3(2)] S[table-format=3.3(2)]}
% \toprule
% \textbf{Process} & {\yfslo~\Mpb} & {\yfsnlo~\Mpb} & {\yfsnnlo~\Mpb} \\
% \hline
% \midrule
% \emup & 246.451(1) & 252.19(1) & 253.05(2) \\
% \emum & 246.452(1) & 251.156(4) & 252.03(5) \\
% \midrule
% \toprule
% \multicolumn{4}{c}{$\left|\pi - \bigl|\Delta \phi_{\mu^+ e^-}\bigr|\right| \leq 3.5~\mrad$} \\
% \hline
% \midrule
% \emup & 196.187(1) & 199.947(1) & 200.01(1)  \\
% \emum & 196.187(1) & 199.608(4)  & 199.682(8) \\
% \toprule
% \end{tabular}
% \caption{Fiducial Cross-Sections for \emupm at different perturbative orders for both scenarios.}
% \end{table}

\begin{figure}
    \centering
    \includegraphics[width=0.45\linewidth]{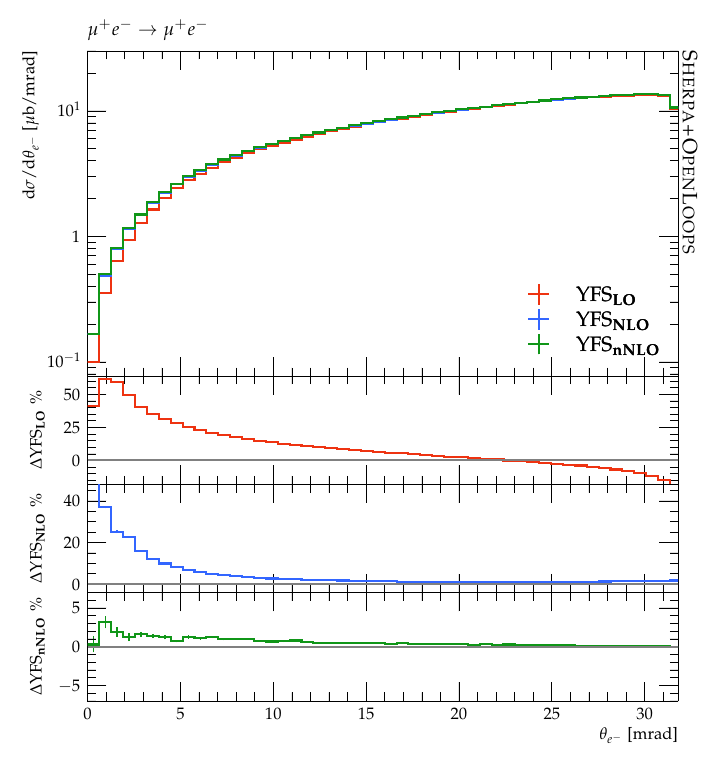}
    \includegraphics[width=0.45\linewidth]{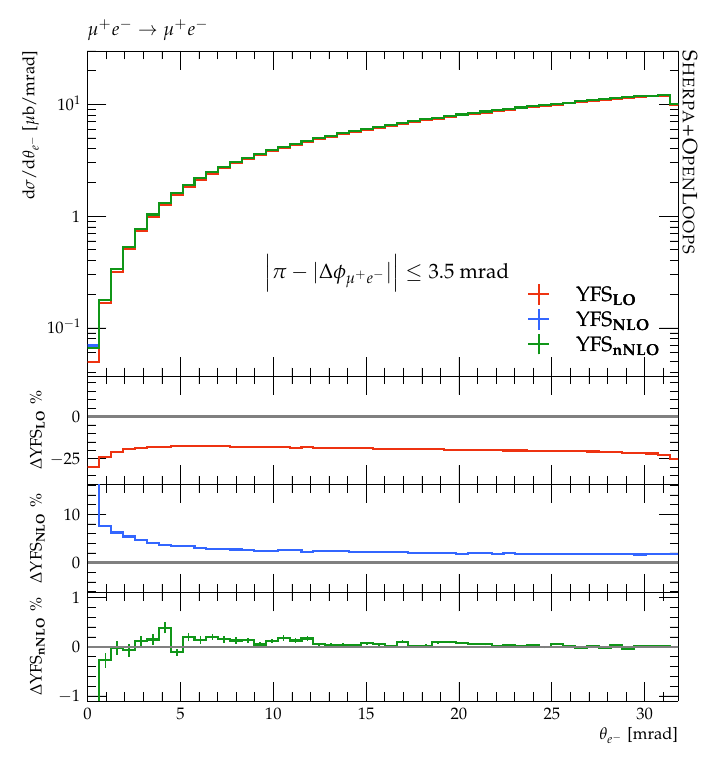}
    \caption{Electron polar angle distribution for different sign scenario for \yfslo(red), \yfsnlo(blue), and \yfsnnlo(green).}
    \label{fig:thetaEDiff}
\end{figure}

In~\cref{fig:thetaEDiff}, we show these three predictions for the polar angle of the electron in the 
lab frame. In the scenario without any acoplanarity cut, we see that \Deltayfs{LO} at large angles has
a negative impact of $~\approx 10\%$, which as we approach $\theta_e=0$ changes signs and grows to be of the order 
of $60\%$. It is interesting to note that this behaviour replicates the NLO predictions of the fixed-order
tools, particularly in the large angle region.
% \footnote{See for example Figure 9. of~\cite{CarloniCalame:2020yoz}}
 This means that \yfslo already captures the dominant effects of the NLO fixed-order predictions, even before we 
include our matching. When the acoplanarity cut
has been included we see that the correction remains fairly flat around 25\%, we attribute the flatness to the reduction
of hard photon radiation, however the YFS will induce a change in the total cross-section hence the overall
normalization. For the 
\Deltayfs{NLO} contributions, we see that the correction becomes flat in the large angle region, around 2\%,
while as we approach the signal region it grows to around 40\%, but with a gentler slope compared to \Deltayfs{LO}. We attribute the size of this correction to the collinear enhanced logarithms which appear for the first time at this order. We see a similar correction in the hard region when employing the acoplanarity
cut, while in the signal region we see a factor 2 reduction when compared to the setup without this cut. 
This implies that a large source of our uncertainty in this region can be attributed to the hard and collinear photon emissions.
For the \yfsnnlo contributions we see quite an improvement in the large angle region. Here the uncertainty due
to missing higher order corrections drops to $~\approx 0.01\%$, which is a similar precision of the LEP era
YFS generators~\cite{Jadach:1995nk,Jadach:1999vf}.

% \begin{table}
%     \centering
%     \begin{tabular}{c|ccc}
%         \emup & LO  & $\text{YFS}_{\textbf{Born}}$ & $\text{YFS}_{\textbf{EEX}}$ \\
%          \hline
%        \Sherpa  & 245.034(3) & 261.296(9)  & 256.315(8) \\
%          \hline
%          \hline
%          & LO & NLO & NNLO \\
%          \hline
%          Mesmer & 245.038910(1) & 255.8437(5) & 256.092(1) \\
         
%     \end{tabular}
    % \begin{tabular}{c|cc}
    %      % \hline
    %    \emup   & LO & NNLO \\
    %      \hline
    %      Mesmer & 245.038910(1) & 256.092(1) \\
    % \end{tabular}
    % \begin{tabular}{c|cc}
    %     \emum & LO  & YFS \\
    %      \hline
    %    \Sherpa & 245.046(8)  & 254.6(1) \\
    %      \hline
    %      \hline
    %      & LO & NNLO \\
    %      \hline
    %      Mesmer & 245.038910(1) & 256.092 \\
    % \end{tabular}
%     \caption{Total cross-sections for \emupm in \Mpb. }
%     \label{tab:xs}
% \end{table}

% \begin{figure}
%     \centering
%     \includegraphics[width=0.45\textwidth]{plots/mesmer-comp/LO/e_t_lab.pdf}
%     \includegraphics[width=0.45\textwidth]{plots/mesmer-comp/LO/muon_t_lab.pdf}
%     \includegraphics[width=0.45\textwidth]{plots/mesmer-comp/LO/muon_theta_lab.pdf}
%     \includegraphics[width=0.45\textwidth]{plots/mesmer-comp/LO/e_theta_lab.pdf}
%     \mycaption{}{LO predictions for the polar angle and Mandelstam variable $t$ for both
%         the muon and electron in the lab frame.}{fig:LO-mesmer}
% \end{figure}

% \subsection{Differential Distributions}

\begin{figure}
    \centering
    \includegraphics[width=0.45\linewidth]{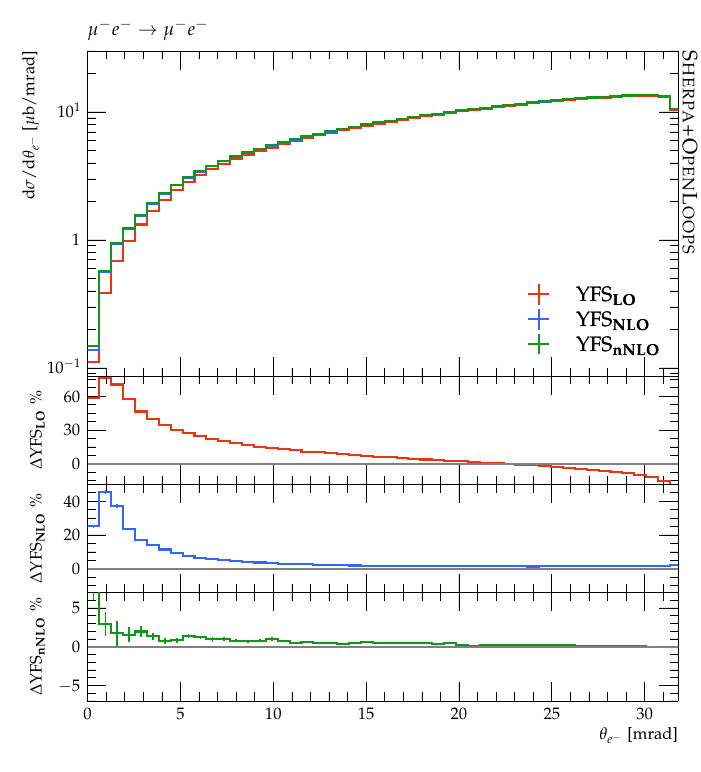}
    \includegraphics[width=0.45\linewidth]{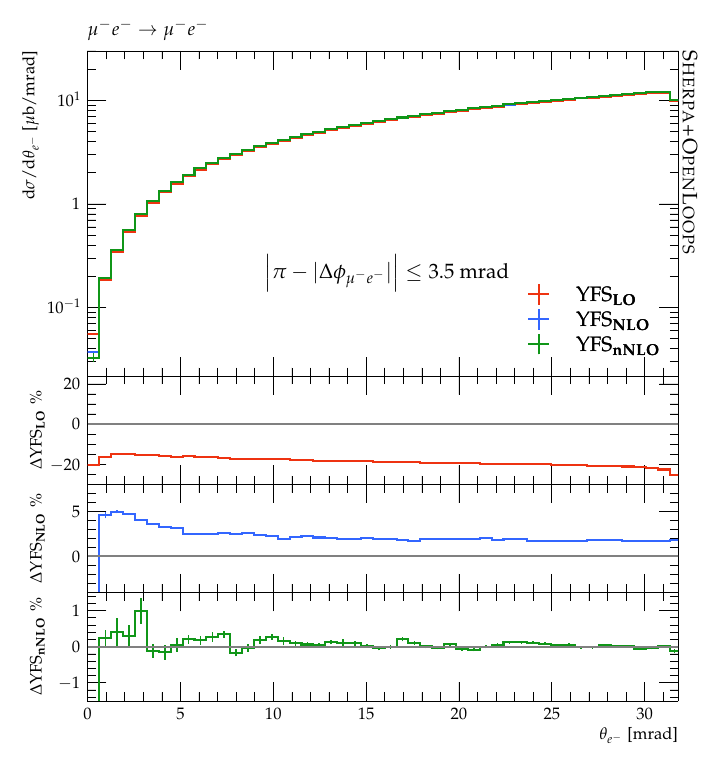}
    \caption{Electron polar angle distribution for same sign scenario for \yfslo(red), \yfsnlo(blue), and \yfsnnlo(green).}
    \label{fig:thetaESame}
\end{figure}

In~\cref{fig:thetaESame}, we present the electron angular distributions for the same-sign scenario. In the first scenario, we see a very similar results as in the \emup case. We note that when the acoplanarity cut is applied we see a smaller  uncertainty for \yfsnlo case.  
\begin{figure}
    \centering
    \includegraphics[width=0.45\linewidth]{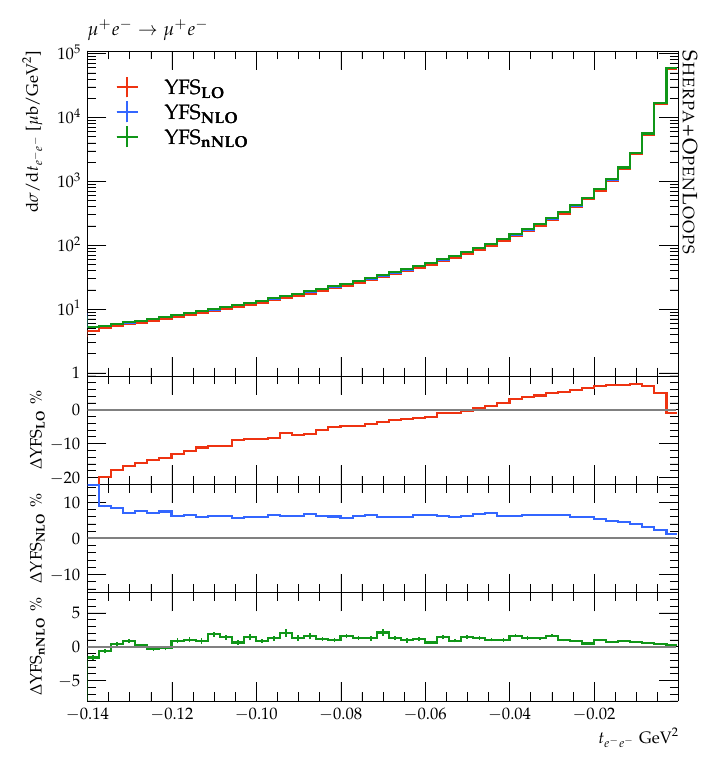}
    \includegraphics[width=0.45\linewidth]{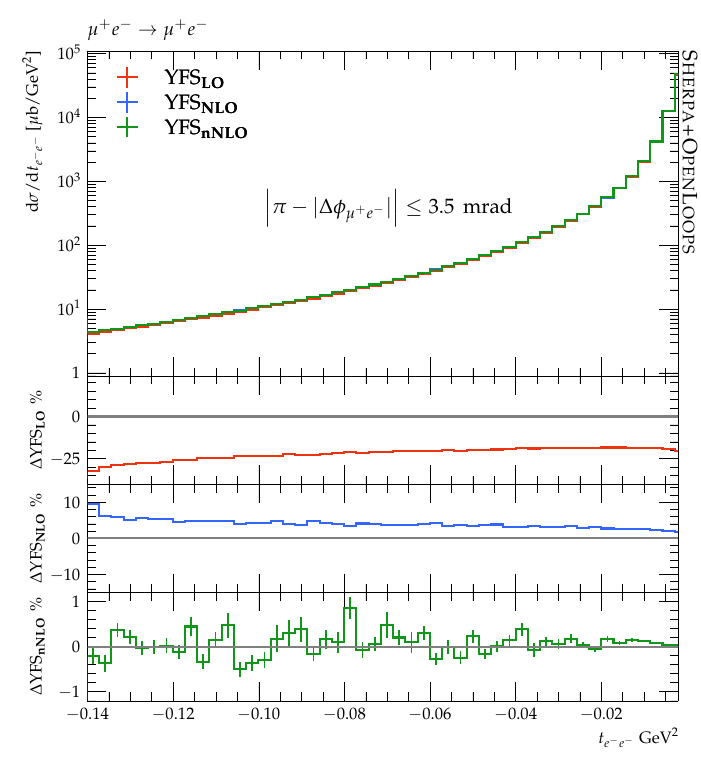}
    \caption{\tee distribution for different sign scenario for \yfslo(red), \yfsnlo(blue), and \yfsnnlo(green).}
    \label{fig:TDiffSign}
\end{figure}
\begin{figure}
    \centering
    \includegraphics[width=0.45\linewidth]{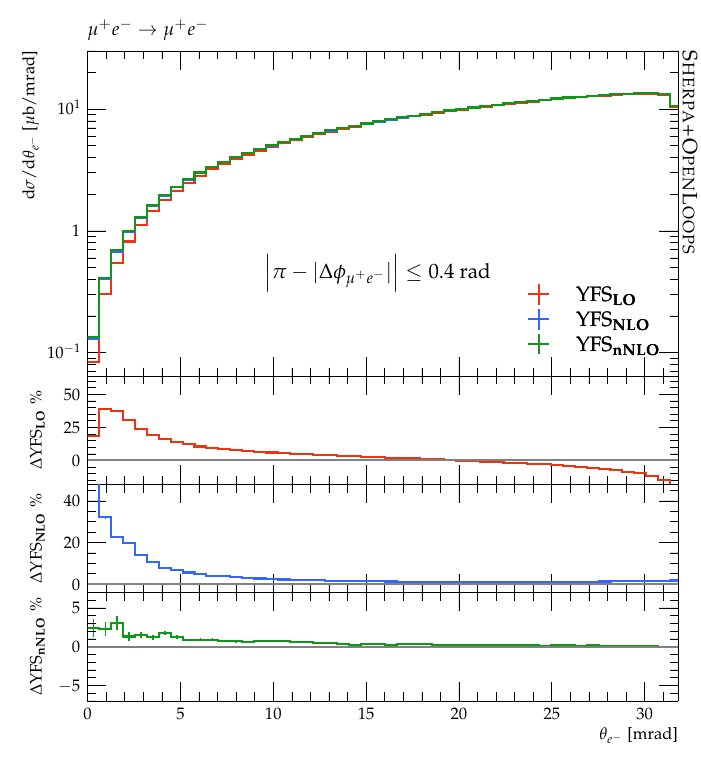}
    \includegraphics[width=0.45\linewidth]{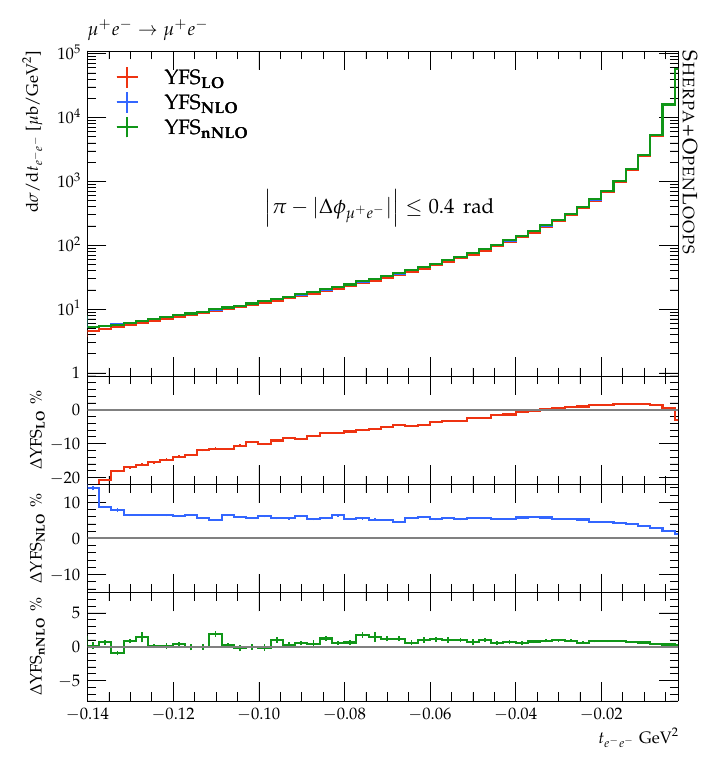}
    \caption{Electron polar angle (left) and \tee (right) distributions for different sign scenarios, including a realistic acoplanarity cut.}
    \label{fig:DiffthetaESetup3}
\end{figure}

\begin{figure}
    \centering
    \includegraphics[width=0.45\linewidth]{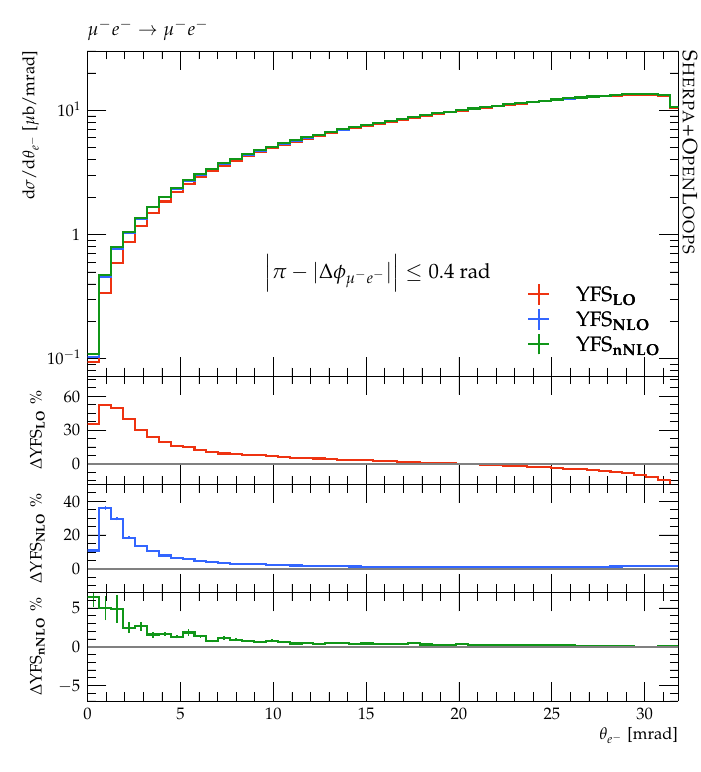}
    \includegraphics[width=0.45\linewidth]{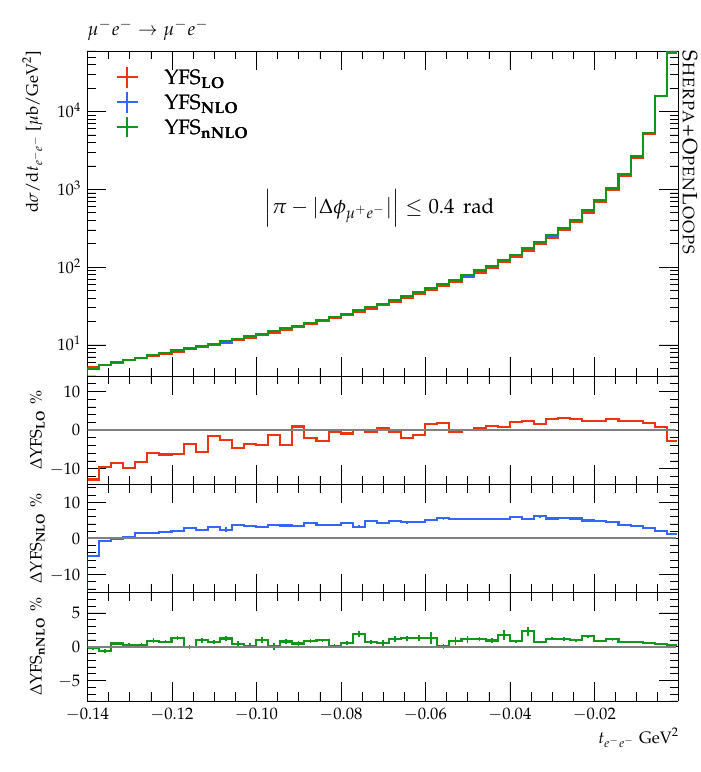}
    \caption{Electron polar angle (left) and \tee (right) distributions for different sign scenarios, including a realistic acoplanarity cut.}
    \label{fig:SamethetaESetup3}
\end{figure}

\Cref{fig:TDiffSign} shows the \tee distribution for \emup. For \yfslo we again see a large correction in the signal region, while in the small $|\tee|$ region we observe a sign change. The shape and the magnitude of $\Deltayfs{LO}$ is similar to that observed by the fixed-order predictions, particularly at low $|\tee|$, while for large $|\tee|$ we have slightly smaller correction. For \yfsnlo we see that our correction becomes relatively flat in the majority of the region while as we approach $\tee=-0.14~\UGeVt$ the effect grows to 
10\%. Finally, for \yfsnnlo we see the corrections remain stable and bring our overall correction down to the percent level. When the additional acoplanarity cut is applied  we see that the \Deltayfs{LO}
is relatively flat around $-25\%$, again reflecting the fact that we have removed a significant amount of the hard ISR/FSR radiation. The \Deltayfs{NLO} is slightly improved over the case without the acoplanarity cut. For \yfsnnlo the corrections are at the per-mille level which is approaching the 10~ppm level, which is roughly factor 10 improvement over the scenario without an acoplanarity cut.  This reflects again that our uncertainty is mostly due to the hard radiation which we have removed with the additional cut.
In~\cref{fig:DiffthetaESetup3,fig:SamethetaESetup3}, we examine the effects of a more realistic acoplanarity cut of 0.4~\rad. We see that \Deltayfs{LO} becomes smaller in the signal region, as the inclusion of this cut suppresses some of the photon radiation generated, which reduces the impact of \yfslo. For the higher-order corrections, the inclusion of this cut has little effect on our uncertainty estimates. This simply reflects the fact that the effect of the cut is present at each order and is largely cancelled in the ratio of~\cref{EQ:Uncert}.  
While we do not have the complete \NNLO correction we tentatively approximate the size of uncertainty by considering the scaling between successive orders as an approximation of missing higher-orders. We can safely do this as we know the YFS series,~\cref{eq:masterYFS}, is well-behaved and free of singularities. In the signal region we can take the approximate missing third order corrections to be of the order $\frac{\Deltayfs{nNLO}}{\Deltayfs{NLO}}\times\Deltayfs{nNLO}$. With this we estimate that our uncertainty in signal region for the first setup is of the order $0.2\%$ while with the addition of the acoplanarity cut this gets further reduced to $0.001\%$ which is within the 10~ppm accuracy needed by MUonE. However, we should stress that this uncertainty arises from missing higher-order corrections and may not reflect the true theoretical uncertainty. In particular, we will need to compare against other approaches, such as collinear-based parton showers, and carefully examine any differences with our YFS approach, incorporating these into our error budget before we can give a conclusive estimate for the complete theoretical uncertainty.

\section{Conclusion}
\label{sec:conclusion}
In this paper, we studied the effect of QED resummation on \emupm process at the proposed MUonE experiment. We found that the effect of this resummation are sizeable particularly in the signal region. From this, we conclude that at the MUonE experiment, one cannot neglect the effects of resummation and it will be mandatory to include if we reach the 10 ppm precision needed ~to extract an accurate measurement of $\Delta\alpha_{\textbf{Had}}$. In particular we have resolved the instabilities seen at $\theta_e < 5~\mrad$ , which had plagued the fixed-order predictions.
We have shown how one can match the YFS resummation to the complete \NLO and \NNLO , where the only approximation we have taken is in regards to the two-loop corrections.
We find that in order-by-order comparisons the uncertainty in our calculation improves, from $\approx 50\%$ for the \yfslo down to $\approx 5\%$ for \yfsnnlo in the inclusive scenario. When apply the additional cut on the acoplanarity these uncertainties enter the per-mil level, approaching the 10~ppm level needed by MUonE experiment. While this acoplanarity cut may be to harsh for the experimental environment, as it may remove too many non-elastic scattering events, it implies that  selectively removing hard radiation events may be one avenue to reaching the precision needed. We estimate that our theoretical uncertainties are at a 
level of $0.2\%$, which is still insufficient for the MUonE experiment, even prior to accounting for method-dependent contributions. It is possible that the inclusion of the $\text{N}^3\LO$ corrections will allow us to reach the precision needs of the MUonE experiment. We believe that it will be feasible to include a subset of these corrections, namely the tree-level and one-loop corrections which will arise at $\text{N}^3\LO$, which we know how to include in the YFS formalism.
However, the two-loop and three-loop contributions pose a significant challenge, and alternative approaches may need to be explored. Another avenue to consider is the potential for combining our YFS-based resummation with a collinear approach. While combining YFS with a full parton-shower presents many difficulties, we might be able to include some higher-order effects arising from the collinear splitting of photons into lepton pairs, as described in~\cite{Flower:2022iew}.

\section*{Acknowledgements}
A.P would like to thank Fred Jegerlehner who reignited my interest in this project and Wieslaw Płaczek for comments on the manuscript and for many discussions on the treatment of YFS form factor. We would also like to thank the MESMER collaboration for multiple conversations and to Carlo Carloni Calame for reading the draft manuscript. 
The work of A.P.\ is supported by grant No. 2023/50/A/ST2/00224 of the National
Science Centre (NCN), Poland.
We gratefully acknowledge Polish high-performance computing infrastructure PLGrid (HPC Center: ACK Cyfronet AGH) for providing computer facilities and support within computational grant no. PLG/2025/018139.
\bibliographystyle{JHEP}
\bibliography{allref}
\end{document}